\documentclass[sigconf]{acmart}
\usepackage{xcolor,soul,framed} 
\usepackage{multirow}
\usepackage{cases}
\usepackage{float}
\usepackage{color}
\usepackage{subfigure}
\usepackage{makecell}
\usepackage{threeparttable}
\usepackage{algorithm}
\usepackage{tcolorbox}
\tcbuselibrary{most}
\usepackage{algorithmic} 
\usepackage{listings}
\usepackage{xcolor}
\usepackage{bm}
\usepackage[normalem]{ulem} 
\renewcommand\footnotetextcopyrightpermission[1]{}
\usepackage{graphicx}  
\usepackage{subfigure}  

\AtBeginDocument{%
  }

\setcopyright{acmlicensed}
\copyrightyear{2018}
\acmYear{2018}
\acmDOI{XXXXXXX.XXXXXXX}

\acmConference[Conference acronym 'XX]{Make sure to enter the correct
  conference title from your rights confirmation emai}{June 03--05,
  2018}{Woodstock, NY}
\acmISBN{978-1-4503-XXXX-X/18/06}

\newtcolorbox{mybox}[2][]
{colback = white, colframe = black,
	colbacktitle = gray, enhanced,
	attach boxed title to top left={yshift=-2mm,xshift = 4mm},
	title=#2,#1}
\begin{document}

\title{Vulnerability-Hunter: An Adaptive Feature Perception Attention Network for Smart Contract Vulnerabilities}


\author{Yizhou Chen}
\email{yizhouchen@stu.pku.edu.cn}
\affiliation{%
  \institution{Peking University}
  \city{Beijing}
  \country{China}
}

\renewcommand{\shortauthors}{Yizhou Chen et al.}

\begin{abstract}
\underline{S}mart \underline{C}ontract \underline{V}ulnerability \underline{D}etection (SCVD) is crucial to guarantee the quality of blockchain-based systems. Graph neural networks have been shown to be effective in learning semantic representations of smart contract code and are commonly adopted by existing deep learning-based SCVD. However, the current methods still have limitations in their utilization of graph sampling or subgraph pooling based on predefined rules for extracting crucial components from structure graphs of smart contract code. These predefined rule-based strategies, typically designed using static rules or heuristics, demonstrate limited adaptability to dynamically adjust extraction strategies according to the structure and content of the graph in heterogeneous topologies of smart contract code. Consequently, these strategies may not possess universal applicability to all smart contracts, potentially leading to false positives or omissions. 
To address these problems, we propose AFPNet, a novel vulnerability detection model equipped with a feature perception module that has dynamic weights for comprehensive scanning of the entire smart contract code and automatic extraction of crucial code snippets (the $P$ snippets with the largest weights). Subsequently, the relationship perception attention module employs an attention mechanism to learn dependencies among these code snippets and detect smart contract vulnerabilities. The efforts made by AFPNet consistently enable the capture of crucial code snippets and enhance the performance of SCVD optimization.
We conduct an evaluation of AFPNet in the several large-scale datasets with vulnerability labels. The experimental results show that our AFPNet significantly outperforms the state-of-the-art approach by 6.38\%-14.02\% in term of F1-score. The results demonstrate the effectiveness of AFPNet in dynamically extracting valuable information and vulnerability detection.
\end{abstract}

\keywords{Quality assurance, Smart contract, Vulnerability detection, Deep learning}

\maketitle

\section{Introduction}
\label{introduction}

Blockchain is essentially an emerging software system that uses smart contracts and computer networks to maintain transaction data. With the emergence and popularity of blockchain and smart contracts, they bind hundreds of millions of virtual assets \cite{roscheisen1998stanford,radziwill2018blockchain,chen2019exploiting,savelyev2017contract}. Increasingly, hackers are attempting to exploit \underline{S}mart \underline{C}ontract \underline{V}ulnerabilities (SCVs) for illicit gain. For example, the YAM Protocol incident in 2020~\cite{werner2021sok} highlighted the risks of unaudited smart contracts and the importance of careful code reviews and testing. In this incident, the YAM protocol, a DeFi protocol built on Ethereum, suffered a critical bug in its smart contract code, resulting in the loss of over \$750,000 worth of investor funds within 35 minutes of the protocol's launch. 

To detect smart contract vulnerabilities (also called smart contract vulnerability detection, abbreviated as SCVD, in the literature, many researchers have put dedicated efforts in designing effective approaches or tools~\cite{luu2016making,tsankov2018securify,mueller2017framework} via using program analysis, fuzzy testing, symbolic execution, etc. These tools rely on expert-defined rules to detect vulnerabilities. However, acquiring expert-defined rules is expensive because these rules are obtained based on expert knowledge and manual summarization. Additionally, considering the increasing number of smart contracts, the existing expert-defined rules can hardly cover all running patterns in smart contracts. Moreover, hackers may easily learn these rules or patterns and bypass them to conduct attacks. 

Currently, an increasing number of deep-learning-based detection methods have been proposed~\cite{zhuang2020smart,liu2021smart,wu2021peculiar,liu2021combining,zhang2022reentrancy}, which achieve higher precision compared with rule-based techniques. 
These approaches introduce graph neural networks to encode the structure graph of smart contract code. However, most of the structure graphs of the smart contracts are sizeable.
Simply encoding the entire structure graph of a program can introduce large computational cost and may not result in satisfactory performance. 
Therefore, existing approaches typically employ strategies that involve simplifying the structure graph of smart contract code through techniques such as graph sampling or graph pooling, following predefined rules. 
The GNNs are then used to encode the simplified graphs and identify SCVs. Unfortunately, the predefined strategies are typically formulated based on static rules or heuristics and may not be fully applicable to structure graphs of smart contract code with heterogeneous topologies, leading to the occurrence of both false positives and false negatives. 
This limitation forces us to explore more intelligent solutions, thereby reducing reliance on predefined rules and devising a methodology for adaptively capturing the vulnerability features of smart contracts.



\begin{figure*}[]
	\centering
	\includegraphics[width=1\textwidth]{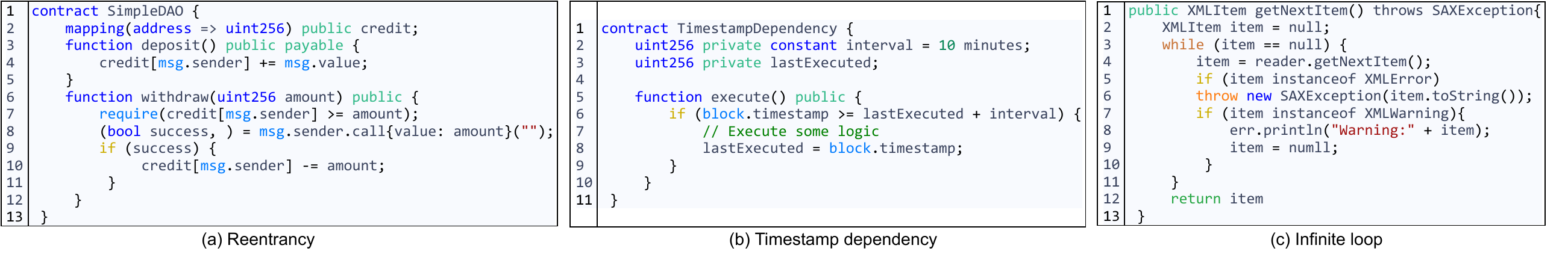}
	\caption{Three examples of smart contract vulnerabilities.}
	\label{fig:scv}
\end{figure*}
To address the above challenges, we propose a \underline{A}daptively \underline{F}eature \underline{P}erception \underline{Net}work (AFPNet). 
Our motivation is that not all elements of a smart contract code possess equal significance. 
AFPNet alleviates the limitations of graph neural networks by focusing on only critical parts of smart contract code that are most likely to trigger vulnerabilities.
Specifically, AFPNet extracts critical vulnerability snippets from smart contract code and encoding these snippets to determine SCVs. 
To realize this idea, AFPNet is consist of two key modules, \underline{F}eature \underline{P}erception \underline{M}odule (FPM) and a \underline{R}elationship \underline{P}erception \underline{A}ttention \underline{M}odule (RPAM). The primary objective of the FPM is to adaptively identify code snippets associated with vulnerabilities and extract them to feed RPAM. Subsequently, the RPAM learns the dependencies of these snippets and predicts the vulnerability.

Specifically, the FPM incorporates $J$ \underline{C}onvolution \underline{N}eural \underline{N}etworks (CNNs) components of varying sizes. It scans the entirety of code sequences in smart contracts, extracting vulnerability feature representations from critical code snippets. These feature representations are abstracted as a series of feature points. The values assigned to these feature points reflect the significance of the corresponding code snippets within the convolution window. The $P$ points with the highest values (indicating crucial information) and an average point with the mean value of a set of feature points (indicating global information) are then selected as crucial features from each convolution window, and finally all the crucial features in the $J$ CNNs are fed to the subsequent RPAM. Subsequently, the RPAM employs a multi-head attention to in-depth interact with these feature information and a fully connected layer is used to determine SCVs. In conclusion, our AFPNet introduces a dynamic feature extraction solution for SCVD, addressing the challenges encountered by existing approaches.

To evaluate the performance of the proposed AFPNet, we conduct extensive experiments on over 40k real-world smart contract instances in two authoritative benchmark datasets and compare it with 15 cutting-edge SCVD methods. According to the experimental results, AFPNet achieves improvements over state-of-the-art methods, i.e., precision, recall, and F1-score improved by an average of 9.93\%, 8.42\%, and 9.12\% for three most common and dangerous vulnerabilities, respectively. Furthermore, we provide theoretical evidence that the time and space complexity of AFPNet are $O(n)$, which has excellent computational efficiency.

In summary, the main contributions of this paper are as follows. 
\begin{itemize}
	\item We propose a novel model called AFPNet, which automates the extraction and interaction of critical vulnerability features for better detecting SCVs.
	\item We have implemented our method, AFPNet, and performed a large-scale evaluation on the three of dangerous SCVs. Quantitative results show that our AFPNet outperforms state-of-the-art SCVD methods and sets the optimum performance.
	\item The source code of AFPNet is publicly available for further research and experimentation. All source code is publicly available at \url{https://anonymous.4open.science/r/AFPNet-461B}.
\end{itemize}

\section{Preliminary}
\label{preliminary}

\subsection{Problem Definition} 

Given a smart contract code snippet $C$ and a deep learning model $M$, our objective is to utilize the model $M$ to predict whether the code snippet $C$ contains SCVs. Let $X$ denote the input space of code snippets and $Y$ represent the set of possible labels, where $Y={0, 1}$ with 0 indicating a non-vulnerable code and 1 indicating a vulnerable code. The underlying vulnerability function is denoted as $f: X \rightarrow Y$. The prediction function learned by the deep learning model, denoted as $h: X \rightarrow Y$, aims to approximate $f$ as accurately as possible. To achieve this, we employ a loss function $\ell$ and train the model $M$ by minimizing the expected loss over a distribution $D$ of code snippets and their corresponding labels. Mathematically, our goal is to minimize the expected loss defined as $\mathbb{E}_{(C, y) \sim D}[\ell(y, h(C))]$.

\subsection{Smart Contract Vulnerability}
Smart contract vulnerabilities~\cite{huang2021hunting} refer to weaknesses or errors in the code of smart contracts that attackers can exploit, leading to manipulation or compromise of the contract's functionality and resulting in financial losses or other negative consequences. This work focuses on three types of the most severe and common SCVs: reentrancy vulnerability, timestamp dependence vulnerability, and infinite loop vulnerability~\cite{zhuang2020smart}. These vulnerabilities pose a significant threat to transaction security on the blockchain, as each of them has the potential to cause substantial financial losses. Fig.~\ref{fig:scv} provides three simple examples of SCVs.

The reentrancy vulnerability allows an attacker to repeatedly enter a function call before the previous invocation of the function is completed, potentially resulting in the execution of malicious code and manipulation of the contract state.
In Part (a) of Fig.~\ref{fig:scv}, the \textit{SimpleDAO} contract keeps track of credits for each user and allows them to withdraw their credits using the \textit{withdraw} function. However, it first sends the requested amount to the caller using the \textit{call} function, and only then updates the credit mapping for the caller. This allows an attacker to repeatedly call the \textit{withdraw} function and execute malicious code within the call function before the credit mapping is updated. An emergency event caused by a reentrancy vulnerability occurred in the DAO hack of 2016~\cite{liu2021combining}, where an attacker exploited a reentrancy vulnerability in the DAO smart contract and stole over \$50 million worth of Ether. 

The behavior of a contract can be affected by the current timestamp, which an attacker may manipulate to gain illegal benefits. This vulnerability is known as timestamp dependence.
In Part (b) of Fig.~\ref{fig:scv}, the \textit{execute} function can be called every \textit{interval} (set to 10 minutes) from the last execution. The contract checks if the current timestamp is greater than or equal to the last execution time plus the interval. If so, it executes some logic and updates the last executed time. However, an attacker can manipulate the block timestamp by mining a block with a timestamp in the future. This can cause the contract to execute multiple times within the \textit{interval}, allowing the attacker to exploit the contract.

Infinite loop vulnerabilities occur when a contract contains a loop that can run indefinitely, consuming excessive amounts of computational resources and causing denial-of-service attack. Part (c) of Fig.~\ref{fig:scv} depicts an example of an infinite loop vulnerability, which can occur in the \textit{getNextItem} function when processing XML streams. In this function, the reader checks for any errors or warnings in the XML stream by calling the \textit{getNextItem} function. However, it is possible that when all XML streams have been read, the reader may always return \textit{null} on a call to \textit{getNextItem}. This failure of the \textit{instanceof} check in the while loop condition will result in the loop executing continuously, always returning \textit{null} from the reader.


\section{Approach}
\label{approach}

This section describes the proposed approach AFPNet in detail. First, we introduce the main idea of FPM and provide a brief overview of the two-block architecture of the AFPNet in Section~\ref{mainidea}. Second, the implementation details of AFPNet are subsequently expounded upon in Sections~\ref{fpm} and~\ref{attention}. Finally, in Section~\ref{complexity}, we theoretically analyze the time and space complexity of AFPNet.

\subsection{Motivation}
\label{mainidea}
\begin{figure*}[]
	\centering
	\includegraphics[width=1\textwidth]{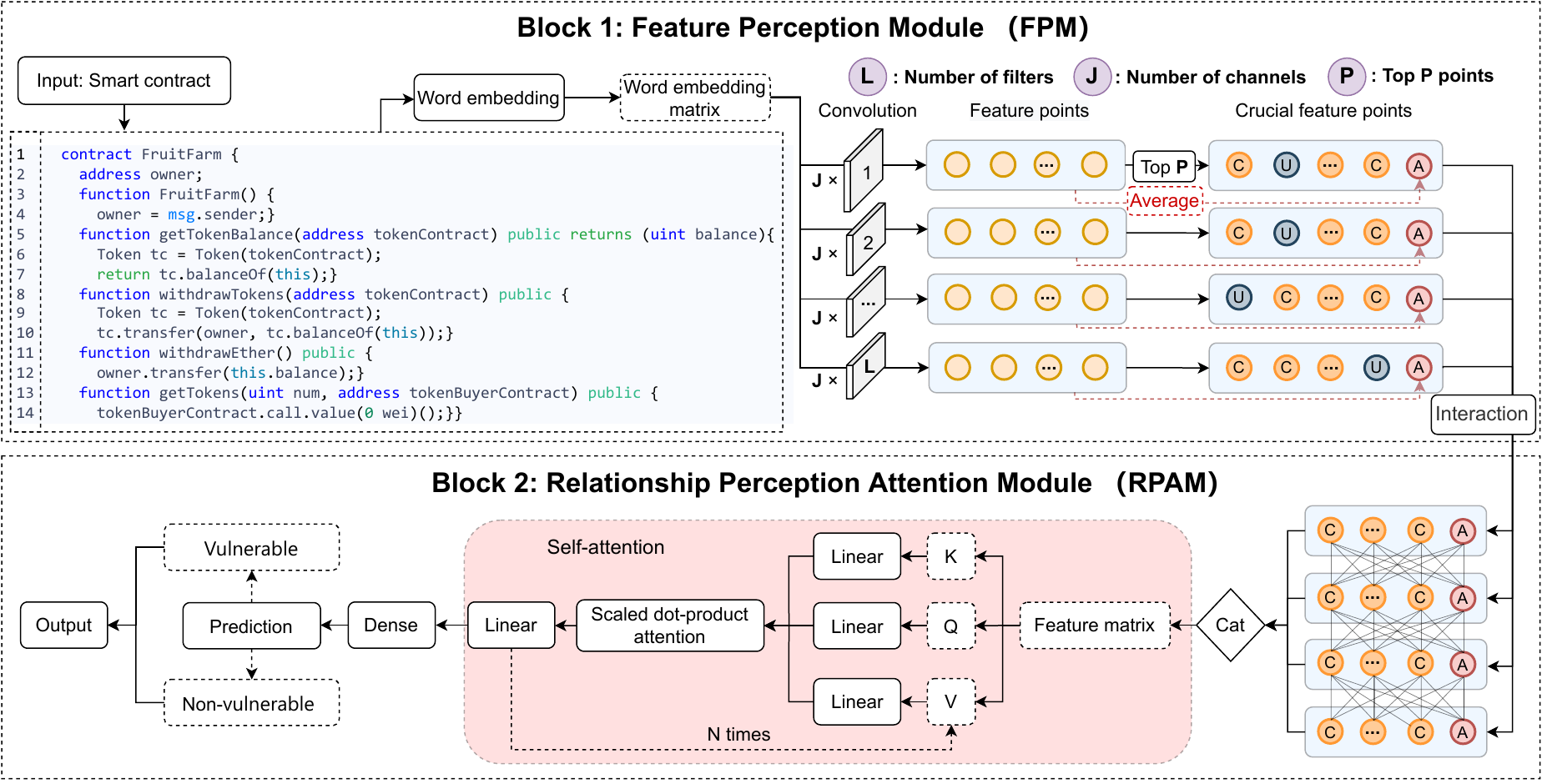}
	\caption{An detailed process of AFPNet. The ``A'' points are average points, indicating the abstraction of the semantic information of the whole code. The ``C" points indicate crucial feature points. The ``U" points indicate unimportant feature points.}
	\label{fig:AFPNet}
\end{figure*}

Our motivation stems from the fact that not all program elements in a smart contract are equally important for detecting vulnerabilities. The top of Fig.~\ref{fig:AFPNet} illustrates a reentrancy vulnerability contract of the real-world, named \textit{FruitFarm}. In the \textit{getTokens} function (line 13 - line 14), the contract invokes a function on the \textit{tokenBuyerContract} address without any checks on the contract's state changes. This means that if the \textit{tokenBuyerContract} invokes a \textit{\textit{getTokens}} function that calls back to the \textit{FruitFarm} contract, it can re-enter the \textit{getTokens} function before the previous invocation completes, allowing for reentrancy attacks. As demonstrated by the above examples, in \textit{FruitFarm} contract, only the \textit{getTokens} function may trigger a vulnerability. In fact, the code snippets that could trigger the vulnerability constitute only a fraction of the entire contract. Previous research \cite{liu2021combining,liu2021smart,zhuang2020smart,wu2021peculiar,zhang2022reentrancy} endeavors have made efforts to extract information regarding potentially vulnerable code from contracts using strategies such as rule-based graph sampling or subgraph pooling. However, the heterogeneous code structure graph of smart contracts poses a significant challenge to the efficacy of these predefined strategies. Hence, our approach is to guide adaptively the deep learning model's focus towards the intended objective, specifically the key program elements associated with the vulnerability. The AFPNet is proposed to realize this idea, as shown in Fig. \ref{fig:AFPNet}. The following subsections provide descriptions of the two key modules of AFPNet, namely FPM and RPAM.

\subsection{Feature Perception Module}
\label{fpm}

This subsection describes the technical details of FPM. The overview of FPM are shown in the first block of Fig.~\ref{fig:AFPNet}. The main goal of FPM is to identify code snippets associated with vulnerabilities and encode them as feature points embedded in a feature matrix. Firstly, we consider a piece of smart contract code as a code sequence, which is sliced into $n$ tokens by a tokenizer, denoted as $CS_{1:n} = \{\bm{t}_1,...,\bm{t}_i,...,\bm{t}_n\}$. $\bm{t}_i$ is the $i$-th token in the $CS_{1:n}$. Then $CS_{1:n}$ is converted to a word embedding matrix $E_{1:n} = \{\bm{e}_1,...,\bm{e}_i,...,\bm{e}_n\}$. $\bm{e}_i \in R^{1,k}$ represents the $k$-dimension word vector corresponding to $\bm{t}_i$. In the following step, we utilize a set of $L$ filters with different heights in the CNNs, which are applied across the entire row in $E_{1:n}$. The objective is to sample features with varying dimensions, thus ensuring a comprehensive and diverse representation of the sampled features. Notably, in order to extract a wide range of informative and non-redundant features, we introduce $J$ convolutional kernels within each filter. These kernels are initialized with unique parameters, promoting the exploration of various local optimal solutions (i.e., focusing on different code snippets) during the training phase. This process is represented as follows: 

\begin{equation}
	C^{j,l} = \mathrm{Relu}(W^{j,l} \bullet E_{i:i + h_l-1} + b),\label{convolution}
\end{equation}
where $W^{j,l} \in R^{h \times k}$ denotes the $j$-th $(j \in J)$ convolution kernel of the $l$-th $(l \in L)$ filter with the height $h_l$. The $C^{j,l} \in R^{1,n-h_l+1}$ represents a new feature maps via a convolution calculation, and $\bm{b}$ is a bias. The $P$ elements with the highest weight $[\widehat{C^{j,l}_1},\widehat{C^{j,l}_2},...,\widehat{C^{j,l}_P}]$ and an average value $A^{j,l}$ from $C^{j,l} \in R^{1,n-h_l+1}$ are concatenated into a new feature vector, which represents approximately the important semantic information of the complete code. FPM generates $J*L$ feature vectors that are concatenated as the output, named $M \in R^{J*L, P+1}$:
\begin{equation} 
	M=
	\left(                 
	\begin{array}{cccccc}   
		\widehat{C}_1^{1,1} & ... & \widehat{C}_p^{1,1} & ... & \widehat{C}_P^{1,1} & \widehat{A}^{1,1}\\ 
		... & ... & ... & ... & ... & ... \\  
		\widehat{C}_1^{j,l} & ... &\widehat{C}_p^{j,l} & ... & \widehat{C}_P^{j,l} & \widehat{A}^{j,l}\\   
		... & ... & ... & ... & ... & ... \\  
		\widehat{C}_1^{J,L} & ... &\widehat{C}_p^{J,L} & ... & \widehat{C}_P^{J,L} & \widehat{A}^{J,L}\\ 
	\end{array}
	\right).
\end{equation}

These sampled feature points provide a condensed representation of the extracted features, enabling subsequent processing to focus on the most important and informative parts of the code. It is noteworthy that the average value $\widehat{A}^{j,l}$ is essential. Indeed, FPM evaluates the importance of the code snippets by the weight of the corresponding feature points. That is to say, we hope that the weight of unimportant feature points decreases and the weight of important feature points increases. Without the average value  $\widehat{A}^{j,l}$, back-propagation can only affect the weight of $P$ feature points. $\widehat{A}^{j,l}$ ensures that the gradient can be conducted to every feature point in $C_{j,l}$ when back propagating. 

\subsection{Relationship Perception Attention Module}
\label{attention}

Within RPAM, we leverage the widely acclaimed and highly effective multi-head attention mechanism to promote information interaction among the feature points in $M$. This facilitates efficient capturing of contextual dependencies within smart contract code. The schematic representation of RPAM can be observed in the second block of Fig. \ref{fig:AFPNet}. The subsequent equation illustrates the mathematical computation of these interdependencies, known as attention scores, to quantify the degree of mutual relevance.


\begin{equation}
	\label{headj}
	head_s=\mathrm{Softmax}(\frac{\bm{q}_s\bm{k}^T_s}{\sqrt{d_k}})\bm{v}_s,
\end{equation}
where the query $q$, key $k$, and value $v$ are the matrices generated by different linear transformations of the input $M$. $head_s$ is the output of the $s$-th attention head.
Subsequently, RPAM builds a fully connected neural network containing two transformation layers and an activation function (ReLU) according to the following equation.

\begin{equation}
	M_C=\mathrm{Concat}(head_1,head_2,\ldots\!,head_S)W,
\end{equation}
\begin{equation}
	M_C'=\max(0,M_C \bullet W_1 + \bm{b}_1)W_2 + \bm{b}_2,
\end{equation}
where $W$, $W_1$ and $W_2$ are weight matrices, $\bm{b}_1$ and $\bm{b}_2$ are bias vectors, and $M_C'$ is the output of multi-head attention. The multi-head attention is repeated $N$ times and outputs a feature matrix.

At the end of the model, we use a fully connected layer and a \textit{sigmoid} function to compute the probability $Y$ in Equation~\ref{sigmoid}. This probability determines whether the input smart contract is vulnerable.

\begin{equation}
	Y = \mathrm{Sigmoid}(FC(M_C')).\label{sigmoid}
\end{equation}

\subsection{Complexity Analysis of AFPNet}
\label{complexity}

Time and space complexity are critical attributes in deep learning models, as they play a vital role in evaluating the efficiency and scalability of the model. In this chapter, we have undertaken a comprehensive analysis and inference of the time and space complexity of AFPNet. To facilitate the analysis of complexity, suppose the input data length is $n$, $K$ is the word embedding size, $S$ and $H$ are the stride and height of the convolution kernel. We set up $J\times L$ convolution kernels in FPM. The multi-head attention uses $h$ attention heads and is repeated $N$ times in RPAM.

\paragraph{Time complexity}
The time complexity is equivalent to the amount of computation, i.e., \underline{FL}oating-point \underline{OP}erations (FLOPs). AFPNet mainly consists of convolution operation $FLOP_{fpm}$ in FPM and attention operation $FLOP_{rpam}$ in RPAM. 

The $FLOP_{fpm}$ is mainly done by the convolution kernel, which computes each time in the input data is $[(n-h)/s + 1] \times k$. The total amount of computation for FPM is as follows:
\begin{equation}
	\begin{aligned}
		FLOP_{fpm} &= [(n-H)/S + 1] \times K \times J \times L \\
		&\approx O(n).
	\end{aligned}
\end{equation}

In RPAM, multi-head attention interacts information by making inner product of any two vectors. The computation of the one-time attention score is as follows:
\begin{equation}
	\begin{aligned}
		FLOP_{rpam} &= (J\times L)^2 \times (P+1) \\
		&\approx O((J \times L)^2).
	\end{aligned}
\end{equation}

As a result, the total amount of computation $FLOP_{gpa}$ of AFPNet is calculated as follows.
\begin{equation}
	\begin{aligned}
		FLOP_{gpa}&= FLOP_{fpm} + N \times FLOP_{rpam} \\
		&= O(n) + N \times O((J \times L)^2)\\
		&\approx O(n).
	\end{aligned}
\end{equation}

\paragraph{Space complexity}
The space complexity of CNN can be expressed model structure and input size. So the space complexity of FPM $Space_{fpm}$ is:
\begin{equation}
	\begin{aligned}
		Space_{fpm} &= O(n \times K \times J \times L \times S \times H) \\
		&\approx O(n).
	\end{aligned}
\end{equation}

According to Section \ref{fpm}, FPM outputs a feature matrix $M \in R^{J \times L, P+1}$ to RPAM. So the space complexity of RPAM $Space_{rpam}$ is:
\begin{equation}
	\begin{aligned}
		Space_{rpam} &= h \times (P+1)^2  + (P+1) \times J \times L \\
		&\approx O(1).
	\end{aligned}
\end{equation}

As a result, the total amount of space complexity $Space_{gpa}$ of AFPNet is calculated as follows.
\begin{equation}
	\begin{aligned}
		Space_{gpa}&= Space_{fpm} + Space_{rpam} \\
		&= O(n) + O(1) \\
		&\approx O(n).
	\end{aligned}
\end{equation}


\section{Experimental Settings}
\label{experimental settings}

\subsection{Research Questions}

\noindent
\textbf{RQ1:} \textit{How effective is AFPNet compared with the state-of-the-art methods in SCVD?}

To answer this question, we test the performance of AFPNet. In addition, to ensure validity, we also compare the AFPNet with 15 state-of-the-art methods in terms of precision, recall, F1-score.

\noindent
\textbf{RQ2:} \textit{How do the each modules of AFPNet contribute to the overall performance?}

To analyze the contribution of each module in AFPNet, We first test the performance of FPM and RPAM separately for validating their capabilities and contribution to the overall effectiveness of AFPNet. Further, we dive deeper to the contribution of FPM and integrate FPM with another three sequence models (i.e., RNN, LSTM, GRU, and Trasnformer) for validating FPM verifying the ability of FPM to adaptively extract features.



\noindent
\textbf{RQ3:} \textit{How does AFPNet work?}

To understand how AFPNet works in SCVD, we manually analyze some example contract cases. We emphasize the code snippets that AFPNet considers most significant and compare the confidence level in AFPNet with that of the state-of-the-art model for the same smart contract.


\subsection{Datasets}
We evaluate the performance of AFPNet on two datasets. 

\textbf{ESC dataset} contains 40,932 Ethereum smart contractsm. Note that the existing ESC dataset are pre-processed smart contracts and not manually labeled datasets. Previous works cited~\cite{liu2021smart,zhuang2020smart} manually labeled a portion of the smart contracts in ESC. We have re-checked the labels of these smart contracts manually to ensure the authenticity of the labels, named ESC. Ultimately, 5013 smart contracts were labeled as reentrancy vulnerability and 4833 contracts were labeled as timestamp dependence vulnerability.

\textbf{VSC dataset} comprises 4,170 smart contracts on the VNT Chain~\cite{zhuang2020smart} that is a blockchain platform similar to Ethereum. In particular, previous work~\cite{liu2021smart,zhuang2020smart} also labeled these data. We likewise re-check all the labels manually and name this dataset as VSC. Ultimately, 237 smart contracts were labeled as infinite loop vulnerability.

In addition, we find that different \textit{.sol} files (a \textit{.sol} file contains one or more smart contracts) in ESC may contain the same smart contract due to reusable transaction logic. That is, the same contract may appear in both the training phase and the testing phase because they belong to different \textit{.sol} files, which is consistent with the distribution of smart contracts in Ethereum. However, this may lead to a misjudgment of the effectiveness of deep-learning-based SCVD methods. This issue seems to have been overlooked by existing researchers. To address this issue, in this paper, we removed duplicate contracts in the ESC, and marked the remaining contracts as ESC$_R$.

\subsection{Baselines}
In the experiment, we use 15 cutting-edge vulnerability detection methods as the baselines. These methods can be classified into rule-based vulnerability detection methods and deep-learning-based vulnerability detection methods.

\textbf{Rule-based vulnerability detection methods} take the code of bycode of a smart contract as input and detect whether the smart contract is vulnerable based on well-defined rules. In this experiment, we include nine representative rule-based vulnerability detection methods, i.e., Smartcheck~\cite{tikhomirov2018smartcheck}, Oyente~\cite{luu2016making}, Mythril~\cite{mueller2017framework}, Securify~\cite{tsankov2018securify}, Slither~\cite{feist2019slither}, Jolt~\cite{carbin2011detecting}, PDA~\cite{ibing2015fixed}, SMT~\cite{kling2012bolt}, and Looper~\cite{burnim2009looper}. 

\textbf{Deep-learning-based vulnerability detection methods} convert the code of a smart contract into a graph for feature learning and then detect whether the contract is vulnerable. In this experiment, we include five state-of-art deep-learning-based vulnerability detection methods, i.e., GCN~\cite{kipf2016semi}, DR-GCN~\cite{zhuang2020smart}, TMP~\cite{zhuang2020smart}, AME~\cite{liu2021smart}, and CGE~\cite{liu2021combining}. 

Additionally, In the domain of software vulnerability detection, there are various general vulnerability detection methods available, such as LineVul~\cite{fu2022linevul}, IVDetect~\cite{li2021vulnerability}, and Devign~\cite{zhou2019devign}. However, none of these methods have yielded satisfactory results specifically for SCVD. Hence, in this paper, we have selected LineVul~\cite{fu2022linevul} as the representative of the general methods due to its superior performance.

For detailed information on all the baseline methods, please refer to Section~\ref{related work}.

\subsection{Evaluation Metrics}
Vulnerability detection is a binary classification task; thus, following existing software-engineering research~\cite{zhao2021comprehensive,zhao2021predicting,zhao2022graph4web,yu2022exploiting}, we measure the performance of a vulnerability detection method via precision, recall, and F1-score metrics. Therefore, we adopt these three metrics in evaluation. In particular, precision measures the ratio of correctly predicted vulnerabilities in all predicted vulnerabilities, recall measures the ratio of correctly predicted vulnerabilities in all vulnerabilities, and F1-score is a harmonic mean between precision and recall. 

\subsection{Implementation Details}
\label{ex_detail}

Following prior works, we randomly select 80\% of the contracts as the training set and the remaining 20\% as the test set for each dataset. The number of training epochs is set to 50. We compute the average result of the last training epoch across five experimental trials as our ultimate result. The experiments are conducted on a server equipped with an Intel i7-10700F CPU (8 cores), 32GiB memory, and two Nvidia RTX 3090 GPUs with a total graphics memory of 48G.

The hyper-parameters of the experiment are set as follows. In the word embedding step, we set the dimensionality of the word embeddings to 256. The FPM component employs five filters of varying window sizes, including filter heights of 2, 3, 5, 7, and 11. Each filter is equipped with 200 internal convolution kernels. The attention layer is recycled 6 times. During training, the learning rate is initialized to 1 x 10$^{-5}$ and is optimized using the AdamW optimizer~\cite{DBLP:journals/corr/abs-1711-05101}. The batch size are fixed at 32. 

\begin{table*}[htbp]
	\centering
	\renewcommand{\arraystretch}{1.3}
	\caption{Performance of studied SCVD methods on the dataset ESC and VSC. ``–'' denotes not applicable.}
	\resizebox{\linewidth}{!}{
		\begin{tabular}{c|c|ccc|ccc|ccc}
			\toprule
			\multirow{2}{*}{\textbf{Row}} & \multirow{2}{*}{\textbf{Methods}} & \multicolumn{3}{c|}{\textbf{RE(ESC)}} & \multicolumn{3}{c|}{\textbf{TD(ESC)}} & \multicolumn{3}{c}{\textbf{IL(VSC)}} \\ \cmidrule{3-8} \cmidrule{9-11} 
			& & \makebox[0.067\textwidth][c]{\textbf{R}} & \makebox[0.067\textwidth][c]{\textbf{P}} & \makebox[0.067\textwidth][c]{\textbf{F}} & \makebox[0.067\textwidth][c]{\textbf{R}} & \makebox[0.067\textwidth][c]{\textbf{P}} & \makebox[0.067\textwidth][c]{\textbf{F}} & \makebox[0.067\textwidth][c]{\textbf{R}} & \makebox[0.067\textwidth][c]{\textbf{P}} & \makebox[0.067\textwidth][c]{\textbf{F}} \\ \midrule
			\textbf{1} & \textbf{Smartcheck} & 32.08 & 25.00 & 28.10 & 37.25 & 39.16 & 38.18 & 23.11 & 38.23 & 28.81 \\
			\textbf{2} & \textbf{Oyente} & 54.71 & 38.16 & 44.96 & 38.44 & 45.16 & 41.53 & 21.73 & 42.96 & 28.26 \\
			\textbf{3} & \textbf{Mythril} & 71.69 & 39.58 & 51.02 & 41.72 & 50.00 & 45.49 & 39.23 & 55.69 & 45.98 \\
			\textbf{4} & \textbf{Securify} & 56.60 & 50.85 & 53.57 & – & – & – & 47.21 & 62.72 & 53.87 \\
			\textbf{5} & \textbf{Slither} & 74.28 & 68.42 & 71.23 & 72.38 & 67.25 & 69.72 & – & – & – \\
			\textbf{6} & \textbf{Jolt} & – & – & – & – & – & – & 23.11 & 38.23 & 28.81 \\
			\textbf{7} & \textbf{PDA} & – & – & – & – & – & – & 21.73 & 42.96 & 28.26 \\
			\textbf{8} & \textbf{SMT} & – & – & – & – & – & – & 39.23 & 55.69 & 45.98 \\
			\textbf{9} & \textbf{Looper} & – & – & – & – & – & – & 47.21 & 62.72 & 53.87 \\
			\midrule
			\textbf{10} & \textbf{GCN} & 78.79 & 70.02 & 74.15 & 75.97 & 68.35 & 71.96  & 63.04 & 59.96 & 64.46 \\
			\textbf{11} & \textbf{DR-GCN} & 80.89 & 72.36 & 76.39 & 78.91 & 71.29 & 74.91  & 67.82 & 64.89 & 66.32 \\
			\textbf{12} & \textbf{TMP} & 82.63 & 74.06 & 78.11 & 83.82 & 75.05 & 79.19  & 74.32 & 73.89 & 74.10 \\
			\textbf{13} & \textbf{AME} & 89.69 & 86.25 & 87.94 & 86.23 & 82.07 & 84.10  & 79.08 & 78.69 & 78.88 \\
			\textbf{14} & \textbf{CGE} & 87.62 & 85.24 & 86.41 & 88.10 & 87.41 & 87.75  & 82.29 & 81.97 & 82.13 \\ \midrule
			\textbf{15} & \textbf{LineVul} & 75.89 & 91.86 & 83.11 & 78.47 & 81.61 & 80.01 & 84.58 & 81.94 & 83.25\\ 
			\midrule
			\textbf{16} & \textbf{AFPNet} & \textbf{95.85} & \textbf{94.60} & \textbf{95.22} & \textbf{93.72} & \textbf{91.00} & \textbf{92.25} & \textbf{93.83} & \textbf{90.32} & \textbf{92.02} \\ \bottomrule
	\end{tabular}}
	\label{tab:rq1_escv}
\end{table*}

\section{Experimental Results}
\label{experimental results}

This section presents the results and analysis of the three research questions.

\subsection{RQ1: How effective is AFPNet compared with the state-of-the-art methods in SCVD?}

\subsubsection{On ESC and VSC}

Table~\ref{tab:rq1_escv} presents the performance of studied vulnerability detection methods on the dataset ESC and VSC, considering three types of most severe and common vulnerabilities (i.e., reentrancy (RE), timestamp dependence (TD), and infinite loop (IL)). And their performance is quantitatively evaluated by Precision (P), Recall (R), and F1-score (F). Note that some vulnerability detection methods can only detect one or two types of vulnerabilities, e.g., Securify~\cite{tsankov2018securify} and Slither~\cite{feist2019slither}, and thus in this table we list only their results of the corresponding vulnerabilities.

We first compare the proposed AFPNet with nine rule-based vulnerability detection methods, i.e., Smartcheck~\cite{tikhomirov2018smartcheck}, Oyente~\cite{luu2016making}, Mythril~\cite{mueller2017framework}, Securify~\cite{tsankov2018securify}, Slither~\cite{feist2019slither}, Jolt~\cite{carbin2011detecting}, PDA~\cite{ibing2015fixed}, SMT~\cite{kling2012bolt}, Looper~\cite{burnim2009looper}. The results of these rule-based vulnerability detection methods are given by the table's top 1-9 rows.

The results show that AFPNet outperforms all existing rule-based  vulnerability detection methods on all three vulnerabilities. AFPNet achieves 33.68\%, 32.31\%, and 70.82\% higher F1-score  compared with the best-performing rule-based methods. Further, we also observe that none of these rule-based vulnerability detection methods can detect three types of vulnerabilities, whereas our proposed AFPNet can.


\begin{table}[htbp]
	\centering
	\renewcommand{\arraystretch}{1.3}
	\caption{The performance of different methods in the AFPNet on the dataset ESC$_R$.}
	\begin{tabular}{c|ccc|ccc}
		\toprule
		\multirow{2}{*}{\textbf{Methods}} & \multicolumn{3}{c|}{\textbf{RE}} & \multicolumn{3}{c}{\textbf{TD}} \\ \cmidrule{2-7} 
		& \makebox[0.03\textwidth][c]{\textbf{R}} & \makebox[0.03\textwidth][c]{\textbf{P}} & \makebox[0.05\textwidth][c]{\textbf{F}} & \makebox[0.03\textwidth][c]{\textbf{R}} & \makebox[0.015\textwidth][c]{\textbf{P}} & \makebox[0.015\textwidth][c]{\textbf{F}} \\ \midrule
		\textbf{TMP} & 78.95 & 78.95 & 78.95 & 76.32 & 78.38 & 77.33 \\
		\textbf{AME} & 78.95 & 81.08 & 80.00 & 78.95 & 76.92 & 77.92 \\
		\textbf{CGE} & 83.61 & 77.20 & 80.27 & 76.20 & 80.44 & 78.77 \\ \midrule
		\textbf{LineVul} & 84.12 & 76.25 & 80.00 & 78.57 & 77.34 & 77.95 \\  
		\midrule
		\textbf{AFPNet} & \textbf{86.05} & \textbf{90.24} & \textbf{88.10} & \textbf{87.80} & \textbf{83.72} & \textbf{85.74} \\ \bottomrule
	\end{tabular}
	\label{tab:rq1_escr}
\end{table}

We then compare AFPNet with six state-of-the-art deep-learning-based detection methods, including GCN~\cite{kipf2016semi}, DR-GCN~\cite{zhuang2020smart}, TMP~\cite{zhuang2020smart}, AME~\cite{liu2021smart}, CGE~\cite{liu2021combining}, and LineVul~\cite{fu2022linevul}. 
The results of these compared methods are given by the 10-15 rows of Table~\ref{tab:rq1_escv}.

The experimental results demonstrate that AFPNet also outperforms the all existing deep-learning-based detection methods on detecting three types of vulnerability. Firstly, when considering all the three performance metrics regarding the three type of vulnerabilities (9 combination cases altogether), AFPNet has the best performance in all cases. To be specific, AFPNet achieves 8.81\%, 4.50\%, and 8.77\% absolute improvement in F1-score over the best baseline method on the three type of vulnerabilities, respectively. The corresponding relative improvements are 10.19\%, 5.12\%, and 10.53\%. Secondly, the general vulnerability detection methods, LineVul, also achieves respectable performance, but our AFPNet still outperforms it by 10.53\%-15.57\% in terms of F1-score.

\subsubsection{On ESC$_R$}

As the ESC dataset contains some overlaps between the training set and test set. Therefore, we remove duplicate contracts as ESC$_R$ dataset and conduct the experiment on this dataset. We select the state-of-the-art AME, TMP, CGE, and LineVul as baselines. This experiment ensures the authenticity of our evaluation and may discover more contracts with vulnerabilities.

Table~\ref{tab:rq1_escr} presents the performance of three deep-learning-based vulnerability detection methods. From this table, we can see that our AFPNet still achieves the best performance among all compared methods in terms of all metrics. The AFPNet outperforms the the best-performing CGE by 2.92\% in terms of recall, by 16.89\% in terms of precision, by 9.75\% in terms of F1-score in reentrancy vulnerability detection; by 15.22\% in terms of recall, by 4.08\% in terms of precision, by 8.85\% in terms of F1-score in timestamp dependence vulnerability detection. This results show the effctiveness of AFPNet.

\subsubsection{PCA dimensionality reduction visualization}

\begin{figure}[]
	\centering
	\includegraphics[width=0.9\linewidth]{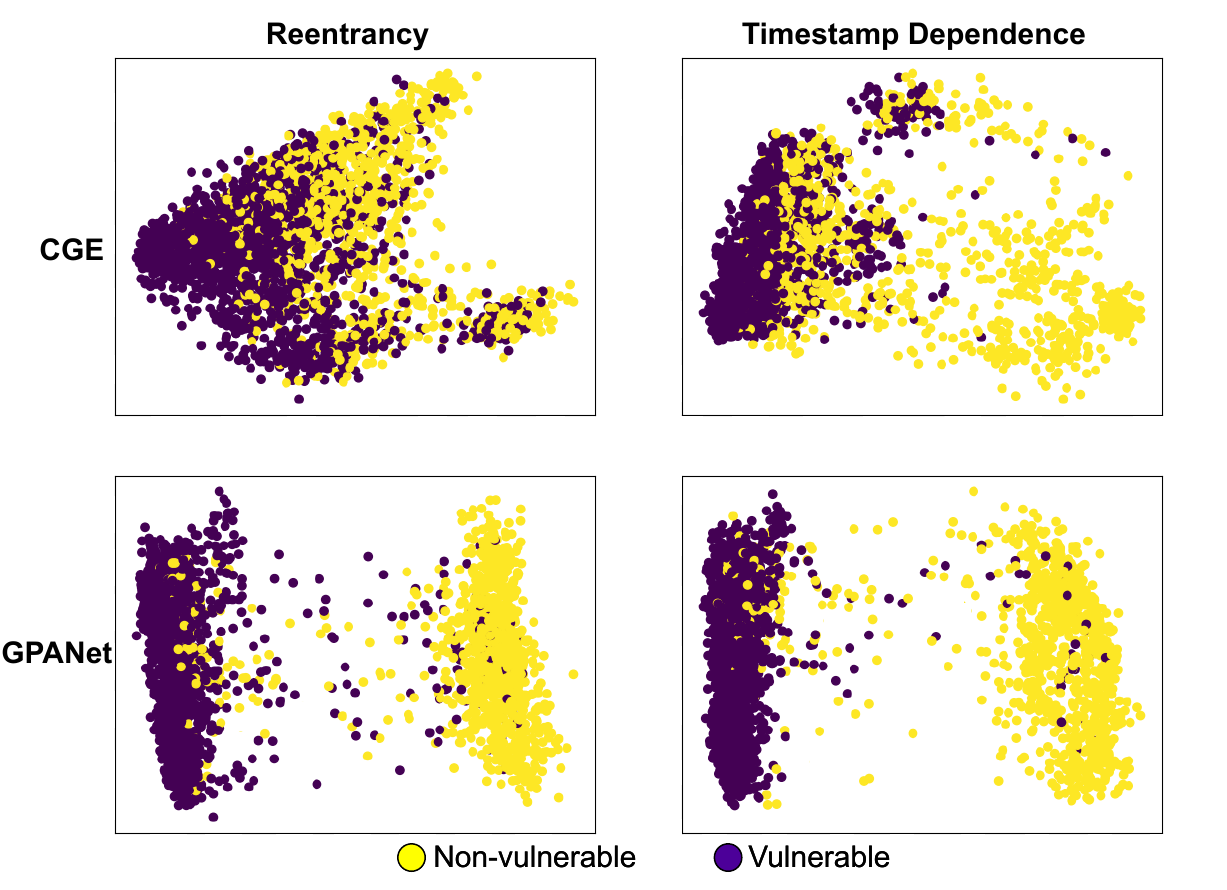}
	\caption{PCA figure, the feature distribution of smart contracts on the ESC$_R$ at  different approaches.}
	\label{fig:FeatureDistribution}
\end{figure}

Using the ESC$_R$ dataset as an example, we employ principal component analysis dimensionality reduction techniques to visualize the feature distribution of AFPNet and the best-performing CGE samples in Fig.\ref{fig:FeatureDistribution}. Each point denotes a contract sample, with purple indicating vulnerability samples and yellow representing non-vulnerability samples. 

In the case of CGE, which includes both vulnerable and non-vulnerable code samples, we observe a considerable overlap between the two categories, making it challenging to distinguish between them. In contrast, AFPNet exhibits a significantly more distinct separation of sample classes. This finding further validates that AFPNet can effectively differentiate between vulnerable and non-vulnerable contracts, enabling accurate detection of security vulnerabilities in smart contract code.

\begin{tcolorbox}
	\textbf{Answer to RQ1}: AFPNet outperforms all the baselines in terms of all metrics. In particular,
	AFPNet achieves 10.19\%, 5.12\%, and 10.53\% improvements in F1-score over the best baseline
	on the ESV and VSC datasets, respectively; achieves 9.75\%, 8.85\% improvements in F1-score over the best baseline on the ESV$_R$ datasets.
\end{tcolorbox}

\begin{table*}[htbp]
	\renewcommand{\arraystretch}{1.3}
	\centering
	\caption{Performance of AFPNet's variants and the traditional sequence vector model on ESC}
	\resizebox{\linewidth}{!}{
		\begin{tabular}{c|c|ccc|ccc|ccc}
			\toprule
			\multirow{2}{*}{\textbf{Row}}  &\multirow{2}{*}{\textbf{Methods}} & \multicolumn{3}{c|}{\textbf{RE}} & \multicolumn{3}{c|}{\textbf{TD}} & \multicolumn{3}{c}{\textbf{IL}} \\ \cmidrule{3-11} 
			& &\makebox[0.076\textwidth][c]{\textbf{R}} &\makebox[0.076\textwidth][c]{\textbf{P}} &\makebox[0.076\textwidth][c]{\textbf{F}} &\makebox[0.076\textwidth][c]{\textbf{R}} &\makebox[0.076\textwidth][c]{\textbf{P}} &\makebox[0.076\textwidth][c]{\textbf{F}} &\makebox[0.076\textwidth][c]{\textbf{R}} &\makebox[0.076\textwidth][c]{\textbf{P}} &\makebox[0.076\textwidth][c]{\textbf{F}} \\ \midrule
			\textbf{1}& \textbf{w/o RPAM} & 83.61 & 77.20 & 80.28 & 74.20 & 83.94 & 78.77 & 78.72 & 76.29 & 77.49 \\
			\textbf{2}& \textbf{w/o FPM } & 59.12 & 80.05 & 68.01 & 63.23 & 87.75 & 73.50 & 67.02 & 70.39 & 68.66 \\ \midrule
			\textbf{3}& \textbf{RNN} & 58.78 & 49.82 & 50.71 & 44.59 & 51.91 & 45.62 & 47.86 & 42.10 & 44.79 \\
			\textbf{4}& \textbf{LSTM} & 67.82 & 51.65 & 58.64 & 59.23 & 50.32 & 54.41 & 57.26 & 44.07 & 49.80 \\
			\textbf{5}& \textbf{GRU} & 71.30 & 53.10 & 60.87 & 59.91 & 49.41 & 54.15 & 50.42 & 45.00 & 47.55 \\ 
            \textbf{6} & \textbf{Transformer} & 65.87 & 75.11 & 70.19 & 75.00 & 73.54 & 74.26 & 59.52 & 84.74 & 69.93 \\
            \midrule
			\textbf{7}& \textbf{FPM-RNN} & 95.44 & 86.79 & 90.91 & \textbf{97.60} & 81.74 & 88.97 & 92.55 & 87.87 & 90.15 \\
			\textbf{8}& \textbf{FPM-LSTM} & \textbf{95.85} & 85.87 & 90.59 & 97.00 & 81.78 & 88.74 & 89.36 & 90.32 & 89.83 \\
			\textbf{9}& \textbf{FPM-GRU} & 95.43 & 88.29 & 91.72 & 92.40 & 89.36 & 90.86 & 90.95 & \textbf{92.43} & 91.68 \\ \midrule
			\textbf{10}& \textbf{AFPNet} & \textbf{95.85} & \textbf{94.60} & \textbf{95.22} & 93.72 & \textbf{91.00} & \textbf{92.25} & \textbf{93.83} & 90.32 & \textbf{92.02} \\ \bottomrule
	\end{tabular}}
	\label{tab:rq2}
\end{table*}

\subsection{RQ2: How do the each modules of AFPNet contribute to the overall performance?}

The development of our AFPNet followed an incremental approach, wherein we started with a basic baseline and progressively added useful components. However, the ablation test was carried out in an inverse fashion, commencing with the full model and systematically removing or substituting individual components with sensible alternatives. This methodology allowed us to investigate the impact of each component on the overall performance of the model and assess their individual contributions to the final results. The results of the ablation tests are reported in Table~\ref{tab:rq2}. Specifically, by ablating RPAM (Row 1), we observe a decrease in the absolute F1-scores of 14.94\%, 13.48\%, and 23.36\% respectively. This finding highlights the significant role played by RPAM in effectively interacting with crucial vulnerability features in SCVD. From another perspective, FPM demonstrates its efficacy in capturing vulnerability features from smart contract code, as evidenced by achieving F1-scores of 80.28\%, 78.77\%, and 77.49\% for the three types of vulnerabilities. Subsequently, by ablating FPM (Row 2), we find that the absolute performance of the F1-score decreases by 27.21\%, 18.75\%, and 23.36\%, respectively, in the three types of vulnerabilities, compared to AFPNet. This is due to that FPM can capture crucial vulnerability features from the smart contract code as analyzed. Without FPM, RPAM can only use the information of the entire code, resulting in a deterioration in performance.


Second, we integrate the proposed FPM with another four sequence models, i.e., RNN, LSTM, GRU, and Transformer (Rows 3-6), and evaluate the integrated models, which are denoted as ``FPM-models'' (Rows 7-10). For example, we use FPM-RNN to represent the RNN model integrated with FPM. Note that in this study, we do not include ``FPM-Transformer'' because RPAM contains a  Transformer similar structure. 
The two studies are performed on the datasets ESC and VSC. The experiment settings are the same as described in Section~\ref{ex_detail}. The results are shown in Table~\ref{tab:rq2}. Due to video memory limitation, when using these sequence models, code length longer than 2000 is intercepted. The ``FPM-model'' does not have this restriction on input length. This is due to that AFPNet has a lower memory cost, which is capable of accommodating longer code sequences in comparison to the other sequence models. 

From the table we can see that the traditional sequence models (i.e., RNN, LSTM, and GRU) do not perform satisfactorily in the three types of vulnerabilities. However, their vulnerability detection performance improves when being inserted with FPM. 
In particular, in reentrancy vulnerability detection, recall, precision and F1-scores improve by an average of 49.91\%, 56.22\%, and 56.35\%, respectively. In timestamp dependence vulnerability detection, recall, precision, and F1-scores improve by an average of 69.68\%, 51.13\%, and 62.85\%, respectively. In infinite vulnerability detection, recall, precision, and F1-scores improve by an average of 64.59\%, 63.12\%, and 77.11\%, respectively. 

Traditional sequence models do not perform well in SCVD tasks, and we suspect the following reasons. (1) A smart contract code sequence is usually very long. Truncating the code sequence may cause serious information loss. (2) The code snippet that triggers vulnerabilities is only a small portion of the entire code. Traditional models are unable to accurately focus on important code snippets. These two reasons result in the low performance. In comparison, our proposed AFPNet substantially alleviates this deficiency and improves the performance of these models significantly. The empirical results show the generalizability of FPM for SCVD.

\begin{tcolorbox}
	\textbf{Answer to RQ2}: FPM contributes significantly to the performance of AFPNet. In combination with RPAM, has resulted in a average F1-score improvement of 18.16\%. Furthermore, compared to all traditional models, the improvement in F1-scores when combined with FPM amounted to 65.44\%.
\end{tcolorbox}

\begin{figure*}[htbp]
	\centering
	\includegraphics[width=1\textwidth]{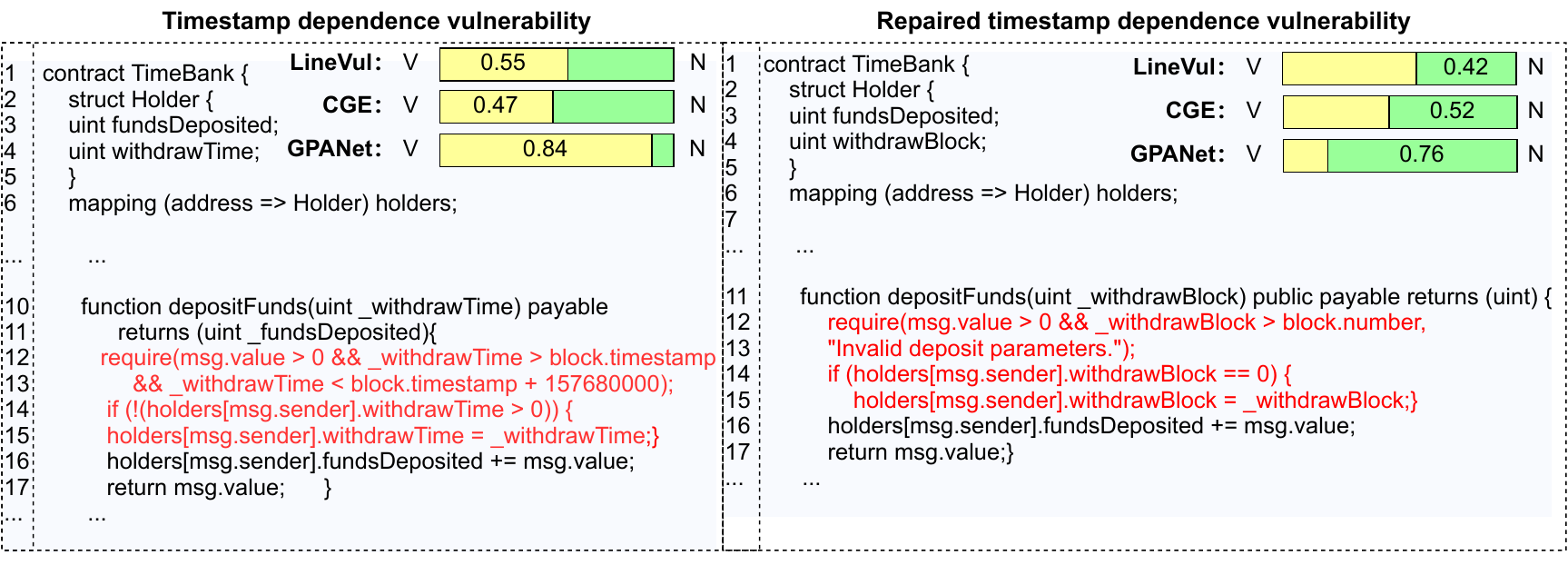}
	\caption{Illustrating example of the real-world smart contracts and their confidence levels through the application of CGE and AFPNet methods.}
	\label{fig:rq3_1}
\end{figure*}

\subsection{RQ3: How does AFPNet work?} 

In this section, we manually analyze several cases where AFPNet works and fails to work, so as to learn how AFPNet works.

\begin{figure}[]
	\centering
	\includegraphics[width=0.9\linewidth]{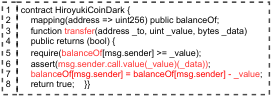}
	\caption{An example of a reentrancy vulnerability contract that is not correctly identified.}
	\label{fig:rq3_2}
\end{figure}

According to Section~\ref{fpm}, the FPM component can sample crucial feature points that correspond to the code snippets in the smart contract code. Therefore, we feed the some smart contract code instances illustrated in the Fig.~\ref{fig:rq3_1} and Fig.~\ref{fig:rq3_2} into the AFPNet that has been trained. The code snippets corresponding to the crucial feature points in the FPM component are extracted and irrelevant punctuation is removed. The words with the highest frequency are marked in red. We anticipate that FPM will consistently prioritize the identification of code snippets that have the potential to trigger vulnerabilities. 

In the left side of Fig.~\ref{fig:rq3_1}, the \textit{TimeBank} contract is designed to implement a straightforward time-locked funding mechanism. However, a significant issue with this code is its failure to adequately ensure that \textit{\_withdrawTime} exceeds the current timestamp, potentially resulting in unexpected behavior until a future point in time. On the right side of Fig.~\ref{fig:rq3_1}, we manually rectify this issue by enhancing security through the introduction of a block number \textit{block.number} denoting the block from which the user is permitted to withdraw funds. Although there is only a minor discrepancy between these two contracts in their conditional statements (line12 - 15), our evaluation, conducted utilizing AFPNet and state-of-the-art CGE, reveals noteworthy findings. In the left contract, the CGE method exhibits a confidence level of 0.47 in identifying vulnerability, while the confidence level for the non-vulnerability is 0.53. This indicates that, the CGE itself exhibits a degree of uncertainty regarding the contract's vulnerability. A similar phenomenon is observed in the case of the right side of the figure. In contrast, our method not only successfully identifies both contracts but also provides a higher confidence level. We attribute this phenomenon to AFPNet's distinctive feature extraction mechanism. Undoubtedly, the red words are the crucial code snippets that are most likely to trigger vulnerabilities. 

But, sometimes AFPNet does not identify vulnerabilities correctly, resulting in a small number of false positives. For example, the Fig.~\ref{fig:rq3_2} shows an example of a reentrancy vulnerability contract that is not correctly identified. The contract is called \textit{HiroyukiCoinDark} and it contains a mapping called \textit{balanceOf} which maps addresses to their corresponding token balance. The contract also contains a function called \textit{transfer} which allows users to transfer their tokens to another address. The vulnerability stems from the use of the \textit{call} function in the \textit{transfer} function. The \textit{call} function is used to call an external function and execute its code within the context of the current contract. In this contract, the \textit{assert} statement after the \textit{call} function assumes that the external function will complete successfully and will not call back into the current contract. An attacker can exploit this assumption by creating a malicious contract. Specifically, the attacker's contract could call the \textit{transfer} function and then immediately call back into the current contract before the transfer is complete. This would allow the attacker to execute arbitrary code within the context of the current contract and potentially modify the contract's state in unexpected ways. Such an obscure reentrancy vulnerability is not common in our dataset, so it is one of the few contracts that has not been predicted successfully. Admittedly, our AFPNet still focuses on key code snippets. 

\begin{tcolorbox}
	\textbf{Answer to RQ3}: the AFPNet can focus on crucial code snippets to identify SCVs. And compared to state-of-the-art models, AFPNet can give higher confidence to improve the performance of SCVD.
\end{tcolorbox}

\section{Discussion}
\label{discussion}

\subsection{How does AFPNet compare with other SCVD methods regarding actual running efficiency?}

We have analyzed the time complexity of AFPNet in Section~\ref{complexity}. But in practice, other factors impact the efficiency (e.g., compilation). Therefore, we discuss the actual running efficiency of AFPNet here. 

Previous work transforms smart contracts to a more advanced form before detecting vulnerabilities through compilation, with three ways in general: (1) compiling the smart contract into data flow graph and control flow graph \cite{liu2021combining,liu2021smart,zhuang2020smart,mueller2017framework,feist2019slither,tsankov2018securify}, (2) compiling smart contracts into bytecode \cite{luu2016making,carbin2011detecting}, and (3) compiling smart contracts into \underline{I}ntermediate \underline{R}epresentation (IR). For example, SmartCheck \cite{tikhomirov2018smartcheck} compiled contract code into extensible markup language and then checked it against the XPath pattern. Looper \cite{burnim2009looper} used an LLVM \cite{lattner2004llvm} compiler to compile the contract code into LLVM IR and then analyzed it and detected vulnerabilities using static analysis techniques. 

From the perspective of detection methods, the rule-based vulnerability detection methods cover all three of the above compilation ways. Deep-learning-based vulnerability detection methods all use the first compilation way. These compilation processes influence the efficiency of SCVD to some extent. 

We randomly select 100 contracts and report their compilation and detection times. We observe that most contracts can be compiled within 5 seconds, while a few intricate contracts may require 10 seconds or even longer. When combined with detection methods, rule-based vulnerability detection methods take about 20 to 60 seconds to compile and  detect a smart contract. In contrast, deep-learning-based vulnerability detection methods require approximately 2 to 5 seconds to compile and detect a smart contract.
Compared to these methods, AFPNet can work directly on the smart contract code, bypassing the cumbersome compilation process. In particular, AFPNet takes less than 0.1 seconds to detect a smart contract. It is worth noting that hardware devices and software versions significantly influence the speed of compilation and detection. The results presented above are based solely on our experimental devices.

\subsection{How does our AFPNet perform compared with some detection methods that focus only on the reentrancy vulnerability?}

Besides the studied detection methods used in the experiment, there are some detection methods for reentrancy vulnerability only. We do not include these methods in the experiment due to generalizability gaps between these approaches and the ones discussed in Section~\ref{experimental results} but present the comparison results in this section. In particular, Wu et al.~\cite{wu2021peculiar} and Zhang et al.~\cite{zhang2022reentrancy} introduced graph-based pre-training techniques to detect reentrancy vulnerabilities with encouraging results on the SmartBugs Wild Dataset~\cite{ferreira2020smartbugs}, which is the largest dataset of reentrancy vulnerabilities with trusted labels available to date. For ease of comparison, we compare the proposed AFPNet with these two techniques (i.e., Peculiar~\cite{wu2021peculiar} and ReVulDL~\cite{zhang2022reentrancy}) on the same dataset. All the parameters are the same as Section~\ref{ex_detail}. The comparison results are listed in Table~\ref{tab:discussion}.

\begin{table}[htbp]
	\renewcommand{\arraystretch}{1.3}
	\centering
	\caption{Comparison results of AFPNet and detection methods focused on reentrancy vulnerability.}
	\begin{tabular}{c|ccc}
		\toprule
		\multirow{2}{*}{\textbf{Methods}} & \multicolumn{3}{c}{\textbf{RE}} \\ \cmidrule{2-4} 
		& \makebox[0.1\textwidth][c]{\textbf{R}} & \makebox[0.1\textwidth][c]{\textbf{P}} & \makebox[0.1\textwidth][c]{\textbf{F}} \\ \midrule
		\textbf{Peculiar} & 92.40 & 91.80 & 92.10 \\
		\textbf{ReVulDL} & 93.00 & \textbf{92.00} & 93.00 \\ \midrule
		\textbf{AFPNet} & \textbf{96.22} & 90.82 & \textbf{93.44} \\ \bottomrule
	\end{tabular}
	\label{tab:discussion}
\end{table}

According to this table, the proposed AFPNet still achieves satisfactory performance and achieves the optimal value with two metrics, i.e., F1-score of 93.44\% and recall of 96.22\%, while the state-of-the-art method, ReVulDL, achieves an F1-score of 93.00\% and recall of 93.00\%. The AFPNet outperforms the ReVulDL by 3.46\% in terms of recall, by 0.47\% in terms of F1-score. Although our AFPNet may have slightly lower precision compared with ReVulDL. But, it is crucial to note that failure to detect vulnerabilities can result in significant financial losses. The cost of misjudging a vulnerable contract far outweighs the cost of misjudging a non-vulnerable one. Therefore, in the task of SCVD, recall is more important compared with precision. In conclusion, AFPNet gives better results without using the pre-trained model and exhibits promising potential in comparison to these specific techniques.

\subsection{How does the widely popular large language model (Chat-GPT) perform in the task of SCVD?}

In previous studies, ChatGPT has demonstrated its competence in comprehending and generating code, exhibiting commendable coding proficiency in diverse coding challenges, including HumanEval and LeetCode, where it has yielded noteworthy results comparable to human performance \cite{bubeck2023sparks,chang2023survey}.
Building upon these findings, our aim is to explore whether the same model can be effectively employed for other tasks, such as vulnerability detection.  In this context, we present an evaluation of the ChatGPT model's performance within the domain of vulnerability detection for smart contract code. 
We employ the OpenAI-provided API to send requests to the Chat-GPT model for the purpose of vulnerability detection. The primary objective of this report is to thoroughly assess the efficacy of utilizing ChatGPT for conducting vulnerability detection tasks and to evaluate its performance using the ESC dataset. The prompt to Chat-GPT is: \textit{Please analyze the following code segment for <type> vulnerabilities. The code segment starts with @. @ <code>}. Where \textit{<type>} is the vulnerability type, and \textit{<code>} is the code that is tested. And every response given by ChatGPT has a clear prediction of whether the smart contract is vulnerability or not.

\begin{table*}[htbp]
	\renewcommand{\arraystretch}{1.3}
	\centering
	\caption{Performance testing of ChatGPT in the SCVD tasks.}
	\begin{tabular}{c|ccc|ccc}
    \hline
    \multirow{2}{*}{\textbf{Vulnerability}} & \multicolumn{3}{c|}{\textbf{ChatGPT}} & \multicolumn{3}{c}{\textbf{AFPNet}} \\ \cmidrule{2-7} 
     & \makebox[0.1\textwidth][c]{\textbf{R}} & \makebox[0.1\textwidth][c]{\textbf{P}} & \makebox[0.1\textwidth][c]{\textbf{F}} & \makebox[0.1\textwidth][c]{\textbf{R}} & \makebox[0.1\textwidth][c]{\textbf{P}} & \makebox[0.1\textwidth][c]{\textbf{F}} \\ \midrule
    \textbf{RE} & 52.77 & 66.50 & 58.84 & 95.85 & 94.60 & \textbf{95.22} \\ \hline
    \textbf{TD} & 46.40 & 69.04 & 55.50 & 93.72 & 91.00 & \textbf{92.25} \\ \hline
    \textbf{IL} & 41.22 & 61.26 & 49.28 & 93.83 & 90.32 & \textbf{92.02} \\ \bottomrule
	\end{tabular}
	\label{tab:discussion_3}
\end{table*}

As show in Table~\ref{tab:discussion_3}, ChatGPT only achieve F1-scores ranging from 49.28\% to 58.84\% in the SCVD task, which is significantly underperforming compared to the state-of-the-art methods. We believe that ChatGPT's limitations in SCVD stem from the following factor. The SCVD is a task heavily reliant on training data to enable the model to accurately identify vulnerabilities. Currently, large language models are frequently trained on generic data, possibly lacking an adequate number of vulnerability examples. This limitation hampers their performance in this area. Compared to our approach, ChatGPT exhibits an absolute performance gap in F1-scores of 36.38\%, 36.75\%, and 42.47\% when detecting three types of vulnerabilities, as compared to AFPNet. Although ChatGPT has strong language and code analysis capabilities, but our AFPNet is better in the SCVD.


\section{Threats of Validity}

\textbf{The threats to external validity} come from the datasets and studied vulnerabilities. To reduce the former threat, we use two publicly available datasets, each consisting of smart contracts marked as vulnerability or non-vulnerability. To reduce the latter threat, we evaluate the studied vulnerability detection methods in three types of most severe and common vulnerabilities.

\textbf{The threats to internal validity} come from the implementation of AFPNet and the compared vulnerability detection methods. To mitigate these threats, we implement AFPNet based on PyTorch and off-the-shelf third-party libraries and adopt the reproducible package of the compared methods. 

\textbf{The threats to construct validity} come from the metrics used to measure the performance of studied vulnerability detection methods. To reduce these threats, we use precision, recall, and F1-score as previous work did~\cite{zhao2021comprehensive,zhao2021predicting,zhao2022graph4web,yu2022exploiting} since vulnerability detection can be regarded as a dichotomous task.

\section{Related Work}
\label{related work}

Smart contract vulnerability detection is regarded as a vital task for blockchain security. In the early work, researchers used several techniques to verify SCVs, including formal verification, symbolic execution, program analysis, and fuzz testing. 
For example, Mueller et al.~\cite{mueller2017framework} designed a comprehensive smart contract analysis tool, Mythril, that integrated static analysis, dynamic analysis, and symbolic execution techniques to detect vulnerabilities and provide analysis reports accurately.
SmartCheck~\cite{tikhomirov2018smartcheck} was a static program analysis tool for finding vulnerabilities, which converted Solidity source code to XML and checked for vulnerabilities based on XPath patterns.
Luu et al.~\cite{luu2016making} used symbolic execution tools to verify SCVs, which received the bytecode of the smart contract and the current global state of Ethereum as input while returning the problem path to the user.
Feist et al.~\cite{feist2019slither} proposed a static analysis framework. It converted smart contract code into an internal representation language, SlithlR, which was then fed into a vulnerability detector to detect contract vulnerabilities. 
Torres et al.~\cite{torres2019art} proposed a vulnerability detection tool, Honeybadger, that used symbolic execution and explicit definitions. It provided the first analysis of honeypot smart contracts and accurately found a large number of honeypot contracts in the real world. 
Mossberg et al.~\cite{mossberg2019manticore} proposed an analytics framework based on dynamic symbolic execution techniques for smart contracts called Manticore. It allowed user-defined analysis and supported both traditional and exotic execution environments. 
Torres et al.~\cite{torres2018osiris} used a framework, Osiris, that combined symbolic execution techniques with taint analysis techniques for accurately detecting integer errors in smart contracts. It detected a wider range of errors while providing better detection specificity. 
Jiang et al.~\cite{jiang2018contractfuzzer} was a fuzz testing tool that generated syntactically compliant input by analyzing the ABI interface of a smart contract. At the same time, new test criteria were defined for different types of vulnerabilities. The EVM monitored the execution status of smart contracts to detect vulnerabilities. 

In recent years, researchers proposed deep learning methods to improve the detection accuracy of SCVs. 
Specifically, Qian et al.~\cite{qian2020towards} proposed snippet representation of smart contracts to extract important semantic information and utilized Bi-LSTM with Attention to detect reentrancy vulnerabilities. 
Zhuang et al.~\cite{zhuang2020smart} represented the semantic structure of the functions of the smart contract via contract graphs and used the GNNs to detect SCVs. 
Liu et al.~\cite{liu2021smart} proposed a vulnerability detection method that combines deep learning with expert rules. This method converted the code into a semantic graph. Then, the graph features and expert rules were fused to verify SCVs. 
Liu et al.~\cite{liu2021combining} encoded expert knowledge as numerical features and then converted the source code into semantic graphs to capture deep graph features. The expert knowledge and graph features are combined to conduct SCVD by GNN.
Wu et al.~\cite{wu2021peculiar} proposed a novel approach, Peculiar, which used a pre-trained technique for detecting the reentrancy vulnerability based on data flow graph. The experiments were conducted on a large dataset, and the results showed that the Peculiar achieved promising performance. 
Zhang et al.~\cite{zhang2022reentrancy} proposed a method named ReVulDL for reentrancy vulnerabilities. It used a graph-based pre-training model to detect reentrancy vulnerabilities and utilized interpretable machine learning to locate the suspicious statements in smart contract.

We also explore general vulnerability detection techniques. The Devign model, proposed by Zhou et al. \cite{zhou2019devign}, represents a notable approach grounded in generalized graph neural networks for the purpose of identifying C++ program vulnerabilities. Li et al. \cite{li2021vulnerability} introduced IVDetect, that harnesses deep learning methodologies to model program dependency graphs, facilitating the identification of vulnerabilities. Fu et al. \cite{fu2022linevul} proposed LineVul, a Transformer-based framework tailored for pinpointing vulnerabilities within C/C++ programs at a granular line-by-line level. By leveraging pre-trained models to capture intricate code semantics, LineVul has emerged as the preeminent method, exhibiting exceptional effectiveness and performance.

In contrast to the aforementioned approaches, the AFPNet possesses the unique capability to adaptive extract crucial vulnerability features from smart contract code. This inherent adaptability allows the AFPNet to effectively capture critical features that are directly related to vulnerabilities, thereby significantly enhancing the performance of SCVD. In comparison to the state-of-the-art approaches, AFPNet demonstrates superior performance.

\section{Conclusions and Future Work}
\label{conclusion}
The SCVs can severely compromise the security of transactions within the blockchain ecosystem. However, the existing approaches employ predefined rule-based strategies to simplify the structure graph of smart contract code, which lack adaptability and consequently lead to false positives. To address this challenge, we propose a novel neural network model, AFPNet, which overcomes the limitations of existing approaches by enabling scanning of the entire code snippet, as well as adaptively extraction and intraction of crucial vulnerability feature to accurately determine the presence of vulnerabilities. The experimental results on several large-scale SCV datasets show that the proposed AFPNet outperforms the state-of-the-art detection methods. In future research, our objective is to further validate the performance of AFPNet by increasing the sample size of vulnerable smart contracts and exploring its application in detecting other types of vulnerabilities. Through enhancing SCVD's performance, our work elevates the security and reliability of blockchain-based systems, both critical factors for successful blockchain transactions.



\bibliographystyle{ACM-Reference-Format}
\bibliography{main}

\end{document}